\documentclass[final]{IEEEtran}
\usepackage{amsmath, amssymb}
\usepackage{graphicx}
\usepackage{subfigure}
\usepackage[noadjust]{cite}
\newtheorem{remark}{\textbf{Remark}}
\usepackage{color}

\title{Toward Simulation-free Estimation of Critical Clearing Time}

\author{Thanh Long Vu, \textit{Member, IEEE}, Surour Al Araifi, \textit{Student Member, IEEE}, Mohamed Elmoursi, \textit{Senior Member, IEEE}, Konstantin~Turitsyn, \textit{Member, IEEE}
\thanks{Thanh Long Vu and Konstantin Turitsyn are with the Department of Mechanical Engineering, Massachusetts Institute of Technology, Cambridge, MA, 02139 USA, e-mail: longvu@mit.edu and turitsyn@mit.edu.
Surour Al Araifi and Mohamed Elmoursi  are with Department of
Electrical Engineering and Computer Science, Masdar Institute, Abu
Dhabi, U.A.E., email: salaraifi@masdar.ac.ae and
melmoursi@masdar.ac.ae.

}}

\begin{document}

\maketitle
\begin{abstract}
Contingency screening for transient stability of large scale,
strongly nonlinear, interconnected  power systems is one of the
most computationally challenging parts of Dynamic Security
Assessment and requires huge resources to perform time-domain
simulations-based assessment. To reduce computational cost of
time-domain simulations, direct energy methods have been
extensively developed. However, these methods, as well as other
existing methods, still rely on time-consuming numerical
integration of the fault-on dynamics. This task is computationally
hard, since possibly thousands of contingencies need to be scanned
and thousands of accompanied fault-on dynamics simulations need to
be performed and stored on a regular basis. In this paper, we
introduce a novel framework to eliminate the need for fault-on
dynamics simulations in contingency screening. This
simulation-free framework is based on bounding the fault-on
dynamics and extending the recently introduced Lyapunov Function
Family approach for transient stability analysis of
structure-preserving model. In turn, a lower bound of the critical
clearing time (CCT) is obtained by solving convex optimization problems
without relying on any time-domain simulations. A comprehensive
analysis is carried out to validate this novel technique on a
number of IEEE test cases.
\end{abstract}


\maketitle

\section{Introduction}


Transient stability assessment, concerned with power systems
stability/instability after contingencies, is a core element of
the Dynamic Security Assessment Systems monitoring and allowing
the reliable operation of power systems around the world. The most
straightforward and dominant approach in industry to this problem
is based on the direct time-domain simulations of transient
post-fault dynamics following possible contingencies. Rapid
advances in computational hardware enable it to perform accurate
simulations of large scale systems possibly faster than real-time
\cite{Huang:2012il,Nagel:2013kf}. However, in practice there are
usually thousands to millions of contingencies that need to be
screened on a regular basis. As such, the computational cost for
time-domain simulations-based transient stability assessment is
huge. At the same time, most of these contingencies are not
critical, and thus most of computational resources are spent for
assessment of contingencies that do not contribute to overall
system risk.

To avoid time-consuming numerical integration of post-fault
dynamics and save the computational resources, the smarter way
nowadays is to use a combination of the direct energy approaches
and time-domain simulation
\cite{Pai:1981dv,chang1995direct,Chiang:2011eo}, in which most
contingencies will be screened by the energy method and the
remaining contingencies are checked by time-domain simulations.
The advantage of direct energy method is that it allows fast
screening of contingencies while providing mathematically rigorous
certificates of stability. After decades of research and
development, the controlling unstable equilibrium point (UEP)
method \cite{Chiang:1994ir} has been widely accepted as the most
successful method among other energy function based direct
screening methods, and is being applied in industry. This method
is based on comparing the post-fault energy with the energy at the
controlling UEP to certify transient stability.

The noticeable drawback of the controlling UEP method is the
inherent difficulty of directly identifying the controlling UEP
\cite{Chiang:2013}. The controlling UEP is defined as the first
UEP whose stable manifold is hit by the fault-on trajectory at the
exit point, i.e. the point where the fault-on trajectory meets the
actual stability boundary of the post-fault Stable Equilibrium
Point (SEP). Note that the actual stability boundary of the SEP is
generally unknown, and thus the computation of the exit point is
very complicated and usually necessitates iterative time-domain
simulations. For a given fault-on trajectory, the controlling UEP
computation requires solving a large set of nonlinear differential
algebraic equations which is done by numerical methods. However,
with respect to these methods, e.g. Newton method, the convergence
region of the controlling UEP can be very small and irregular
compared to that of the SEP. If an initial guess for the numerical
solver was not sufficiently close to the controlling UEP, then the
computational algorithm will result in wrong controlling UEP and
might probably converge to a SEP, leading to unreliable stability
assessment. Unfortunately, it is extremely hard to find an initial
guess sufficiently close to the controlling UEP.

The second drawback of the controlling UEP method is that it
requires simulating and storing each fault-on trajectory to carry
out the assessment for the respective contingencies. To the best
of our knowledge, there are only a few works on contingency
screening without relying on fault-on dynamics simulations.
Particularly, in \cite{roberts2015algebraic} the closest UEP
method is exploited and an algebraic formulation of the critical
clearing time is obtained based on polynomial approximation of the
swing equations. However it is assumed that the dynamics of the
rotor angles during  the  fault  is  a constant positive
acceleration. This approximation is remarkable and may cause
incorrect estimation of the critical clearing time.

The objective of this paper is to develop novel numerical approach
that can potentially alleviate the computational burden of finding
the controlling UEP. We aim to achieve this objective by
developing a completely simulation-free technique for the
estimation of critical clearing time. This technique is based on
an extension of the recently introduced  Lyapunov Functions Family
(LFF) approach \cite{Vu:2014}. The principle of this approach is
to provide transient stability certificates by constructing a
family of Lyapunov functions and then finding the best suited
function in the family for given initial states. Basically, this
method certifies that the post-fault dynamics is stable if the
fault-cleared state stays within a polytope surrounding the
post-fault equilibrium point and the Lyapunov function at the
fault-cleared state is smaller than the minimum value of Lyapunov
function over the flow-out boundary of that polytope. Therefore,
to screen the contingencies for transient stability, this method
only requires the knowledge of the fault-cleared state, instead of
the whole fault-on trajectory.

Exploiting this advantage of LFF method, a technique is introduced
to bound the fault-on dynamics and thereby the fault-cleared
state. This bound leads to a transient stability certificate that
only relies on checking the clearing time, i.e. if the clearing
time is under certain threshold then the fault-cleared state is
still in the region of attraction of the original SEP and
 the post-fault dynamics
is determined stable. By this new method, a fast transient
stability assessment for a large number of contingencies can be
obtained without using any simulations. Such approach can be
utilized in several power system applications, such as optimal
power flow, resources allocation, and HVDC control problems
\cite{4487649, 6575172,
6887372,6755588,6837516,6194234,6629400,6315377}, where the
proposed transient stability certificate can help reduce the
search space by eliminating less critical contingencies in
studies.

The structure of this paper is as follows. In Section
\ref{sec:LFF} the contingency screening problem addressed in this
paper is introduced, together with the extension of the LFF
approach for transient stability analysis. Section
\ref{sec:certificate} presents the main result of this paper
regarding the simulation-free algebraic estimation of the critical
clearing time, and explains how this new stability certificate can
be used in practice to screen contingency for transient stability
without any time-domain simulations. Finally, in Section
\ref{sec:simulations} performance of the proposed method on
contingency screening of several IEEE test systems is presented
and analyzed. Section \ref{sec:discussion} concludes the paper
with discussions about possible ways to improve the algorithms.

\section{Lyapunov Function Family Approach for Transient Stability}
\label{sec:LFF}

In this section, we show that the Lyapunov function family
approach \cite{Vu:2014}, originally presented for the
Kron-reduction model, is applicable to the transient stability
analysis of structure-preserving power models. Then, we extend
this family to a set of convex Lyapunov functions family, that
will be instrumental to establish a lower bound of critical
clearing time in the next section.

In normal conditions, power grids operate at some stable
equilibrium point. During disturbances such as faults, the system
evolves subject to the fault-on (disturbance) dynamics and moves
away from the pre-fault equilibrium point.  After the fault is
cleared, the system may return back to the pre-fault SEP or to a
new post-fault SEP depending on whether the fault is self-cleared
or cleared by circuit breakers action. In this paper, the proposed
method tackles the type of contingencies, where a fault occurs in
a transmission line and then self clears such that the post-fault
network recovers to the pre-fault network topology. To describe
the  post-fault dynamics, we utilize the differential
structure-preserving model \cite{bergen1981structure}. This model
naturally incorporates the dynamics of rotor angle as well as
response of dynamic load power output to frequency deviation.
Though it does not model the dynamics of voltage in the system, in
comparison to the Kron-reduction models with constant impedance
loads \cite{386159}, the structure of power systems and the impact
of load dynamics are preserved in this approach. When the losses
of the transmission lines are ignored, the model can be expressed
as:
\begin{align}
\label{eq.structure-preserving}
 m_k \ddot{\delta_k} + d_k \dot{\delta_k} + \sum_{j \in
  \mathcal{N}_k} a_{kj} \sin(\delta_k-\delta_j) = &P_{m_k},  \\
  &k=1,\dots,m,  \nonumber\\
  \label{eq.structure-preserving2}
  d_k \dot{\delta_k} + \sum_{j \in
  \mathcal{N}_k} a_{kj} \sin(\delta_k-\delta_j) = &-P^0_{d_k},  \\
  & k=m+1,\dots,n, \nonumber
\end{align}
where the first $m$ equations represent the dynamics of generators
and the remaining $(n-m)$ equations represent the dynamics of
frequency-dependent loads. With $k=1,...,m,$ then $m_k$ is the
dimensionless moment of inertia of the $k^{th}$ generator, $d_k$
is the term representing primary frequency controller action on
the governor, and $P_{m_k}$ is the effective dimensionless
mechanical power input acting on the rotor. With $k=m+1,...,n,$
then $d_k>0$ is the constant frequency coefficient of load and
$P^0_{d_k}$ is the nominal load. Let $\mathcal{E}$ be the set of
all the transmission lines and $\mathcal{N}_k$ be the set of
neighboring buses of the bus $k^{th}.$ Then,
$a_{kj}=V_kV_jB_{kj},$ where $[B_{kj}]_{\{k,j\} \in \mathcal{E}}$
is the  susceptance matrix and
 $V_k$ represents the voltage magnitude at the $k^{th}$ bus, both of which are assumed to be constant. The
stationary operating condition is given by
$[\delta_1^*,\dots,\delta_n^*, 0,\dots,0]^T$ where $\delta_k$ is
solution of the power flow-like equations
\begin{align}\label{eq.operatingCondition}
\sum_{j \in
  \mathcal{N}_k} a_{kj} \sin(\delta_k-\delta_j) =P_k, \forall k=1,\dots,n,
\end{align}
where $P_k=P_{m_k}, k=1,\dots,m,$ and $P_k=-P^0_{d_k},
k=m+1,\dots,n.$ We assume that there exists a stable operating
condition $\delta^* \in \Delta(\lambda), \lambda<\pi/2,$ where the
polytope $\Delta(\lambda)$ is defined by inequalities
$|\delta_{kj}|\le \lambda$ for all $\{k,j\}\in \mathcal{E}.$

In the LFF approach, the nonlinear couplings and the linear model
are separated. To do that, the state vector $x = [x_1,x_2,x_3]^T$
is introduced which is composed of the vector of generator's angle
deviations from equilibrium $x_1 = [\delta_1 - \delta_1^*,\dots,
\delta_m - \delta_m^*]^T$, their angular velocities $x_2 =
[\dot\delta_1,\dots,\dot\delta_m]^T$, and the vector of load's
angle deviation from equilibrium
$x_3=[\delta_{m+1}-\delta_{m+1}^*,\dots,\delta_n-\delta_n^*]^T$.
Let $E$ be the incidence matrix of the corresponding graph, so
that $E[\delta_1\dots\delta_n]^T =
[(\delta_k-\delta_j)_{\{k,j\}\in\mathcal{E}}]^T$. Consider matrix
$C$ such that $Cx=E[\delta_1\dots\delta_n]^T.$  Consider the
vector of nonlinear power flow $F$ in the simple trigonometric
form $
F(Cx)=[(\sin\delta_{kj}-\sin\delta^*_{kj})_{\{k,j\}\in\mathcal{E}}]^T.$

Then, in state space representation the system can be expressed in
the following compact form:
\begin{align}
\dot{x}_1 &=x_2 \nonumber \\
\dot{x}_2 &=M_1^{-1}D_1x_2-S_1D^{-1}E^TSF(Cx)  \\
\dot{x}_3 &= -S_2D^{-1}E^TS F(Cx) \nonumber
\end{align}
where $S=\emph{\emph{diag}}(a_{kj})_{\{k,j\}\in \mathcal{E}}$ is
the diagonal matrix of  coupling magnitudes and $S_1=[I_{m\times
m}\quad O_{m\times n-m}], S_2=[O_{n-m\times m} \quad I_{n-m\times
n-m}], D_1=\emph{\emph{diag}}(d_1,\dots,d_m),
M_1=\emph{\emph{diag}}(m_1,\dots,m_n),
D=\emph{\emph{diag}}(m_1,\dots,m_m,d_{m+1},\dots,d_n).$
Equivalently,
\begin{equation}\label{eq.Bilinear}
 \dot x = A x - B F(C x),
\end{equation}
with the matrices $A,B$ given by the following expression:
\begin{align*}
A=\left[
        \begin{array}{ccccc}
          O_{m \times m} \qquad & I_{m \times m}  \qquad & O_{m \times n-m}\\
          O_{m \times m} \qquad & -M_1^{-1}D_1 \qquad & O_{m \times n-m} \\
          O_{n-m \times m} \qquad &O_{n-m \times m} \qquad & O_{n-m \times n-m}
        \end{array}
      \right],
\end{align*}
and
\begin{align}
\label{eq.Bmatrix}
 B= \left[
        \begin{array}{ccccc}
          O_{m \times |\mathcal{E}|}; \quad
          S_1D^{-1}E^TS; \quad
          S_2D^{-1}E^TS
        \end{array}
      \right].
\end{align}
Here, $|\mathcal{E}|$ is the number of edges in the graph defined
by the susceptance matrix, or equivalently the number of non-zero
non-diagonal entries in $B_{kj}$.

\begin{figure}
\centering
\includegraphics[width = 3.5in]{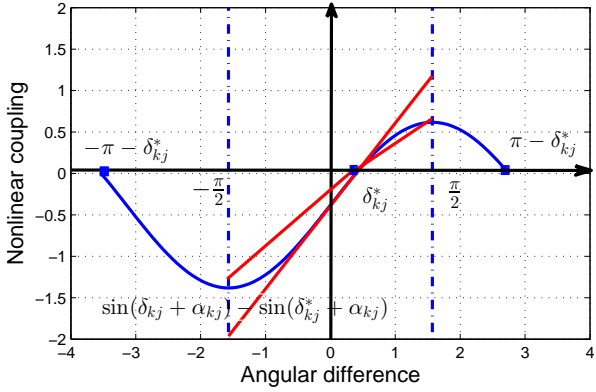}
\caption{Strict bounding of the nonlinear function $f_{kj}$ by
linear functions of the angular difference
$(\delta_{kj}-\delta^*_{kj})$ in the lossy power systems}
\label{fig.StrictBoundingLoss}
\end{figure}

For the system defined by \eqref{eq.Bilinear}, the LFF approach
proposes to use the Lyapunov functions family  given by:
\begin{align} \label{eq.Lyapunov}
V(x) = \frac{1}{2}x^T Q x - \sum_{\{k,j\}\in \mathcal{E}}
K_{\{k,j\}} \left(\cos\delta_{kj}
+\delta_{kj}\sin\delta_{kj}^*\right)
\end{align}
in which  the diagonal, nonnegative matrices $K, H$  and the
symmetric, nonnegative matrix $Q$ satisfy the following linear
matrix inequality (LMI):
\begin{align}
\label{eq.QKH}
    \left[   \begin{array}{ccccc}
          A^TQ+QA  & R \\
          R^T  & -2H\\
        \end{array}\right] &\le 0,
  \end{align}
with $R = QB-C^TH-(KCA)^T$. Then, it can be proved that the
Lyapunov function is decreasing in the polytope $\mathcal{P}$
defined by inequalities $|\delta_{kj}+\delta_{kj}^*| \le \pi,
\forall \{k,j\} \in \mathcal{E}.$ In order to ensure that the
system will not escape the polytope $\mathcal{P}$ during transient
dynamics one condition will be added to restrict the set of
initial states inside $\mathcal{P}.$ Accordingly, we define the
minimization of the function $V(x)$ over the union
$\partial\mathcal{P}^{out}$ of the flow-out boundary segments
$\partial\mathcal{P}_{kj}^{out}$ as follows:
\begin{align}\label{eq.Vmin1}
 V_{\min}=\mathop {\min}\limits_{x \in \partial\mathcal{P}^{out}} V(x),
\end{align}
where $\partial\mathcal{P}_{kj}^{out}$ is the flow-out boundary
segment of polytope $\mathcal{P}$ that is defined, for each
transmission line $\{k,j\}\in \mathcal{E}$ connecting generator
buses $k$ and $j,$ by $|\delta_{kj} +\delta_{kj}^*| = \pi$ and
$\delta_{kj}\dot{\delta}_{kj} \ge 0$. Given the value of
$V_{\min},$ an LFF-based estimation for the region of attraction
of the equilibrium point is given by
\begin{align}\label{eq.invariant}
 \mathcal{R_P} = \left\{x \in\mathcal{P}: V(x) < V_{\min}\right\}.
\end{align}

Finally, to determine if the post-fault dynamics is stable, we
check if the fault-cleared state $x_0$  is inside the stability
region estimate $\mathcal{R_P}$, i.e. if $x_0$ is in the polytope
$\mathcal{P}$ and $V(x_0) < V_{\min}.$ Therefore, to certify
transient stability of each contingency, the LFF approach only
need to know the fault-cleared state $x_0$ (i.e. the state of
fault-on trajectory at the clearing time), rather than the whole
fault-on trajectory.

In this paper, the proposed approach is only concerned with
voltage phase angles staying inside the polytope $\mathcal{Q}$
defined by inequalities $|\delta_{kj}| \le \pi/2, \forall \{k,j\}
\in \mathcal{E}.$ An advantage of considering this polytope of
voltage phasor angles is that inside this polytope the Lyapunov
function $V(x)$ defined in \eqref{eq.Lyapunov} is convex. As such,
the minimum value $V_{\min}$ can be calculated in polynomial time.
In addition, inside this polytope, a stricter bounding for the
nonlinear flow vector $F$ can be established as follows
\begin{align}
\label{eq.NonlinearBounding}
(f_{\{k,j\}}-(\delta_{kj}-\delta_{kj}^*))(f_{\{k,j\}}-\beta(\delta_{kj}-\delta_{kj}^*))
\le 0
\end{align}
where $\beta = \dfrac{1-\sin\lambda}{\pi/2-\lambda}>0$ and
$f_{\{k,j\}}=\sin\delta_{kj}-\sin\delta_{kj}^*$ is an element of
the vector $F.$ Exploiting this strict bound of the nonlinear flow
vector $F,$ the LMI \eqref{eq.QKH} can be replaced by the
following less restrictive LMI:
\begin{align}
\label{eq.NewQKH}
    &\left[   \begin{array}{ccccc}
          A^TQ+QA-2\beta C^THC   & \tilde{R} \\
          \tilde{R}^T  & -2H\\
        \end{array}\right] \le 0, \\
       & \tilde{R}=QB-(1+\beta)C^TH-(KCA)^T, \nonumber
  \end{align}
while all the above results for the stability certificate still
hold true. In particular, the estimate for region of attraction is
given by
\begin{align}\label{eq.RoAestimate}
 \mathcal{R_Q} = \left\{x \in\mathcal{Q}: V(x) < V_{\min}\right\}
\end{align}
with
\begin{align}
\label{eq.Vmin2} V_{\min}=\mathop {\min}\limits_{x \in
\partial\mathcal{Q}^{out}} V(x).
\end{align}
The proof of this fact is given in Appendix \ref{appen.NewQKH}.
With the less restrictive LMI \eqref{eq.NewQKH}, a broader family
of Lyapunov functions can be obtained, which will be exploited to
establish the lower bound of the critical clearing time in the
next section.

\begin{remark} The main drawback of the proposed
stability certificate is that it currently does not incorporate
the detailed model of generators and its associate control
systems, such as excitation systems, PSS and governor system.
Swing equation model doesn't incorporate associated control
systems and generator's fast dynamics and assumes a fixed field
voltage magnitude during transient period. However, the setpoint
values of voltage magnitude can be allowed to fluctuate around the
nominal value $V_0$ (let say less than $10\%$ around $V_0$). In
the matrix $B$ in \eqref{eq.Bmatrix}, we take the new the coupling
magnitude diagonal matrix
$S=\emph{\emph{diag}}(1.1^2V_0^2B_{kj})_{\{k,j\}\in \mathcal{E}}.$
Consider the new nonlinear vector
$F=[f_{kj}]_{\{k,j\}\in\mathcal{E}}$ where
\begin{align}
f_{kj}=\frac{V_kV_j(\sin\delta_{kj}-\sin\delta^*_{kj})}{1.1^2V_0^2}
\end{align}
We can see that the bounding for nonlinear function $f_{kj}$ in
\eqref{eq.NonlinearBounding} still holds true with $\beta$
replaced by the smaller value $ 0.9^2\beta/1.1^2.$ Then, all the
other results will follow accordingly. As such, the simple
Lyapunov function  \eqref{eq.Lyapunov} and stability region
estimate \eqref{eq.RoAestimate} can be easily extended to the case
when voltage magnitude setpoints fluctuate $10\%$ around the
nominal value. In this case, since we have looser bounding for the
nonlinear vector $F,$ the according stability region estimate will
be smaller than the original defined in \eqref{eq.RoAestimate}.
Therefore, the proposed framework can manifest the fact that the
stability region is smaller due to the effects of generators'
control systems (however, it cannot capture the  voltage collapse
phenomenon  when the voltage magnitudes sag to the low values).
From this analysis, we suggest that in the practical transient stability assessment, we should accordingly modify the estimation of the stability region to avoid overestimation of the CCT
due to the usage of simple generators' model.
\end{remark}

\begin{remark} Since the proposed stability
certificate only requires the Lyapunov function to be locally
decreasing, rather than decreasing in the whole state space as in
the energy method, the LFF framework can be extended to
incorporate the losses in transmission lines. Indeed, the
stability analysis here is essentially based on bounding the
nonlinear function $f_{kj}$ by linear functions of $\delta_{kj}$
as in \eqref{eq.NonlinearBounding}, i.e. whenever the bounding
\eqref{eq.NonlinearBounding}  holds true, we can have the
stability region estimate accordingly. For the power systems with
losses, we take the coupling magnitude diagonal matrix
$S=\emph{\emph{diag}}(V_kV_jY_{kj})_{\{k,j\}\in \mathcal{E}}$ and
the nonlinear function $f_{kj}$ as
\begin{align}
f_{kj}=
(\sin(\delta_{kj}+\alpha_{kj})-\sin(\delta^*_{kj}+\alpha_{kj})
\end{align}
Here, $Y_{kj}=\sqrt{G_{kj}^2+B_{kj}^2}$ and
$\alpha_{kj}=\arctan(G_{kj}/B_{kj}) \ll 1,$ where $G_{kj}$ and
$B_{kj}$ are the (normalized) conductance and susceptance of the
transmission line $\{k,j\}.$ From Fig.
\ref{fig.StrictBoundingLoss}, we can show that the nonlinear
bounding \eqref{eq.NonlinearBounding} still holds true for any
$x\in \mathcal{Q}$ and
\begin{align}
\beta=\min_{\{k,j\}\in\mathcal{E}}
\frac{\sin(\pi/2+\alpha_{kj})-\sin(|\delta^*_{kj}|+\alpha_{kj})}{\pi/2-|\delta_{kj}^*|}
\end{align} Then, all the stability analysis follows accordingly. Therefore, the LFF framework and the CCT estimation to be presented in the next section is applicable to lossy power systems. We will illustrate the proposed framework for estimating CCT of the lossy 2-bus system in Section IV.A.
\end{remark}

\section{Contingency Screening without Time-domain Simulations}
\label{sec:certificate}

In this section, we present a new approach to the contingency
screening  problem, which relies on a combination the LFF
framework introduced in the previous section and the bounding for
the reachability set of the fault-on dynamics, through which we
can guarantee that the fault-cleared state is still inside the
region of attraction of the post-fault stable equilibrium point.
Interestingly, this bound leads to an algebraic simulation-free
lower bound of the critical clearing time. Therefore, this
contingency screening approach completely removes any time-domain
simulations of both the post-fault dynamics and fault-on dynamics.

\subsection{Bounding for The Fault-on Dynamics}

If the time-domain simulation for fault-on dynamics is used, the
fault-cleared state $x_0$ can be determined by directly
integrating the fault-on dynamics. Then, the value of $V_0 =
V(x_0)$ computed from \eqref{eq.Lyapunov} is compared to the value
of $V_{\min}$ to certify transient stability.

Now, assume that time-domain simulations are not used to integrate
the fault-on dynamics. Then the fault-cleared state $x_0$ will not
be known precisely. To guarantee that $x_0 \in \mathcal{Q}$ and $V
(x_0) < V_{\min},$ we will bound the fault-on dynamics. Consider
the normal condition when the pre-fault system is in the stable
operating condition defined by the stable equilibrium point
$\delta^*_{pre}\in \Delta(\lambda).$ Assume that a fault occurs at
the transmission line $\{u,v\} \in \mathcal{E}$ and then
self-clears such that the power network recovers to its pre-fault
topology.
During the fault, the power system dynamics is approximated by equations:
\begin{align}
\label{eq.fault-on} \dot{x}_F=Ax_F-BF_{pre}(Cx_F) +
BD_{\{u,v\}}\sin\delta_{{uv}_F}
\end{align}
Here, the fault-on trajectory is denoted as $x_F(t)$ to
differentiate it from the post-fault trajectory $x(t)$ in
\eqref{eq.Bilinear}. $D_{\{u,v\}}$ is the unit vector to extract
the nonlinear function
$(\sin\delta_{{uv}_F}-\sin\delta^*_{{uv}_{pre}})$ from the
nonlinear vector
$F_{pre}=[(\sin\delta_{{kj}_F}-\sin\delta^*_{{kj}_{pre}})]_{\{k,j\}\in\mathcal{E}}$,
which serves to model the elimination of the faulted line
$\{u,v\}$ during the fault. In Appendix
\ref{appendix.BoundingCondition}, the following center result
regarding the bounding of the fault-on dynamics is proven, which
will be instrumental to the introduction of stability certificate
in the next section. If there exist matrices $Q,K,H, H\ge 0$ and a
positive number $\gamma$ such that
\begin{align}
\label{eq.BoundingCondition}
    \left[   \begin{array}{ccccc}
          \tilde{A}+ \gamma (QBD_{\{u,v\}})(QBD_{\{u,v\}})^T  & \tilde{R} \\
          \tilde{R}^T  & -2H\\
        \end{array}\right] &\le 0,
  \end{align}
where $\tilde{A}=A^TQ+QA - 2\beta C^THC,\tilde{R} =
QB-(1+\beta)C^TH-(KCA)^T$, then along the fault-on dynamics
\eqref{eq.fault-on} we have $\dot{V}(x_F(t)) \le
\dfrac{1}{2\gamma}$ whenever $x_F(t)$ being in the polytope
$\mathcal{Q}.$

Note that due to \eqref{eq.BoundingCondition}, the Lyapunov
function's derivative $\dot{V}(x)$ along the post-fault dynamics
\eqref{eq.Bilinear} is non-positive in the polytope $\mathcal{Q}.$
Basically, the above result provides a certificate to make sure
that the fault-on dynamics does not deviate too much from the
post-fault dynamics. As such, if the clearing time is under some
threshold, then the fault-cleared state (i.e. the state of
fault-on system at the clearing time) is not very far from the
considered working condition. The above result as such is
essential to establish a lower bound of the critical clearing time
in the next section.

\subsection{Estimation of The Critical Clearing Time}
\label{sec.EstimationCCT}

Let the clearing time be $\tau_{clearing}.$ In Appendix
\ref{appendix.ClearingTimeCertificate}, the following stability
certificate which only relies on checking the clearing time is
proven. If the inequality \eqref{eq.BoundingCondition} holds and
the clearing time $\tau_{clearing}$ satisfies
$\tau_{clearing}<2\gamma (V_{\min}-V(x_{pre})),$ where
$x_{pre}=\delta^*_{pre}-\delta^*_{post},$  then, the fault-cleared
state $x_F(\tau_{clearing})$ is still inside the region of
attraction of the post-fault SEP $\delta^*_{post}$ and the
post-fault dynamics following the considered contingency leads to
the stable operating condition $\delta^*_{post}$.

Therefore, this stability certificate provides us with a lower
bound of the critical clearing time as $2\gamma
(V_{\min}-V(x_{pre}))$ obtained by solving the inequality
\eqref{eq.BoundingCondition}. This estimation is totally
simulation-free,  distinguishing it from other methods in the
literature to estimate the critical clearing time.

\begin{figure}[t!]
\centering
\includegraphics[width = 2.6in]{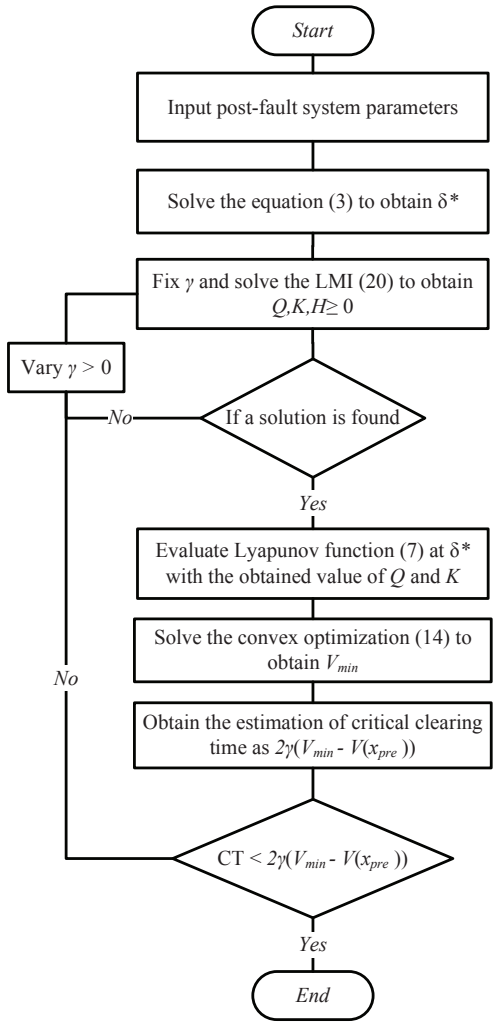}
\caption{Algorithm to screen contingencies for transient stability
without simulations of fault-on dynamics and post-fault
dynamics}\label{flowchart}
\end{figure}

We note that it is also possible to extend this stability
certificate to the case when several contingencies co-exist. This
case is of practical interest. Indeed, the large-area blackout in
practice is usually  a result of multiple contingencies happening
at short time interval. Though large-area blackout is rare, its
effect is severe, both economically and humanly. Therefore, it is
critical  to check if the power grids stand when several
contingencies are happening, or leading to large-area blackout.
The technique presented in this paper provides a framework to
certify the safety of power grids.

\subsection{Choosing Lyapunov Function and
Parameter $\gamma$} Since there is a family of Lyapunov functions
$V(x),$ characterized by matrices $Q,K,$ and positive numbers
$\gamma$ that satisfy the inequality \eqref{eq.BoundingCondition},
we have different estimations $2\gamma (V_{\min}-V(x_{pre}))$ of
the critical clearing time (CCT). To get the highest possible
estimation of the CCT, we need to find the maximum value of
$2\gamma (V_{\min}-V(x_{pre}))$ over all the matrices $Q,K$ and
positive numbers $\gamma$ satisfying \eqref{eq.BoundingCondition}.
Unfortunately, this is an NP-hard, strongly nonlinear optimization
problem with both nonlinear objective function and nonlinear
constraint.

We observe that a good selection of Lyapunov function and the
parameter $\gamma$ is obtained if we can predict the location of
the fault-cleared state. In the following, we propose two
procedures  suggesting some directions to search for feasible
Lyapunov function and parameter $\gamma$ allowing for good
estimation of the CCT. The first procedure is totally heuristic,
where we vary $\gamma$ and find the corresponding Lyapunov
function. The second one is based on a prediction of the
fault-cleared state. Both of these procedures rely on solving a
number of convex optimization problems in the form of either
quadratic programming or semidefinite programming.

\textbf{Procedure 1:} To solve the inequality
\eqref{eq.BoundingCondition}, we note that for a fixed value of
$\gamma,$ the inequality \eqref{eq.BoundingCondition} can be
transformed to the following LMI of the matrices $Q,K,H$ via Schur
complement:
\begin{align}
\label{eq.BoundingConditionLMI}
    \left[   \begin{array}{ccccc}
          A^TQ+QA-2\beta C^THC   & (\sqrt{\gamma}(QBD_{\{u,v\}}) \;\; \tilde{R}) \\
          (\sqrt{\gamma}(QBD_{\{u,v\}}) \;\;\tilde{R})^T  & -L\\
        \end{array}\right] &\le 0,
  \end{align}
  where $L=\left[   \begin{array}{ccccc}
          I   & O \\
          O  & 2H\\
        \end{array}\right].$
The matrices $Q,K,H$ can be found quickly from the LMI
\eqref{eq.BoundingConditionLMI} by convex optimization. Therefore,
a heuristic algorithm can be used to find solution of
\eqref{eq.BoundingCondition}, in which $\gamma$ is varied and the
LMI \eqref{eq.BoundingConditionLMI} is solved to obtain the
matrices $Q,K,H$ accordingly.

\textbf{Procedure 2:}
\begin{itemize}
\item [1)] Calculate the distance $r$ from the equilibrium point $\delta^*_{post}$ to the boundary of the polytope $\mathcal{Q}$ as $r=\min_{\delta \in \partial \mathcal{Q}}||\delta-\delta^*_{post}||_2.$
\item [2)] Take $k$ points $x_1,...,x_k$ uniformly distributed on the sphere $S=\{\delta: ||\delta-\delta^*_{post}||_2=r\}$ which surrounds $\delta^*_{post}$ and stays inside $\mathcal{Q}.$ These points are considered as possible predictions for the fault-cleared state.
\item [3)] For each point $x_i,$ using the adaptation algorithm proposed in \cite{Vu:2014} to find a Lyapunov function $V_i(x)$ characterized by matrices $Q_i,K_i$ such that the point $x_i$ stays inside the stability region estimate $\mathcal{R_Q}$ defined in \eqref{eq.RoAestimate}. This adaptation algorithm can quickly find a suitable Lyapunov function after a finite number of steps.
\item [4)] For the matrices $Q_i,K_i,$ find the maximum value $\gamma^*_i$ satisfying the inequality \eqref{eq.BoundingCondition} as: $\gamma^*_i=\max \gamma$ subject to \eqref{eq.BoundingCondition} where $Q=Q_i,K=K_i,H=H_i$. Calculate $\tau_i =2\gamma^*_i (V_{{\min}_i}-V_i(x_{pre})).$
\item [5)] Take the estimation of the CCT as the maximum value out of $\tau_1,...,\tau_k.$
\end{itemize}

We note that compared to Procedure 1, Procedure 2 may remarkably
increase the computational complexity of calculating the CCT
estimate. Recent studies shown that matrices appearing in power
system context are characterized by graphs with low maximal clique
order, and thus the related SDP in these procedures can be quickly
solved by the new generation of SDP solvers
\cite{Javadmadani2014sdp, Jabr2012}. In addition, the advances in
parallel computing, e.g. distributed computing with zero overhead
communication, promises to significantly reduce the computational
load for these SDP solvers.

\subsection{Contingency Screening without Simulations}

The stability certificate in Section III.B provides us with a way
to directly screen contingencies for transient stability
assessment without any time-domain simulations, as described by
the algorithm in Fig. \ref{flowchart}. Basically, for the
contingency manifested by the tripping of line $\{u,v\},$ one can
check if the inequality \eqref{eq.BoundingCondition} is solvable.
In case it is solvable to find the matrices $Q,K,H,$ and the
positive number $\gamma,$ then the Lyapunov function $V(x)$ can be
derived as in \eqref{eq.Lyapunov}, and the minimum value
$V_{\min}$ defined in \eqref{eq.Vmin2} can be calculated. Finally,
if the clearing time (CT) $\tau_{clearing}$ satisfies that
$\tau_{clearing}<2\gamma (V_{\min}-V(x_{pre})),$ where
$x_{pre}=\delta^*_{pre}-\delta^*_{post},$ then we conclude that
the post-fault dynamics following the considered contingency leads
to a stable operating condition. If this inequality is not true,
or if there is no solution for the inequality
\eqref{eq.BoundingCondition}, then nothing can be concluded about
the stability or instability of the post-fault dynamics. The
contingency in this case should be screened by other energy method
or by direct time-domain simulations.



In contingency screening, it is greatly advantageous if we have a
certificate to screen any possible contingency associated with the
tripping of any transmission line in the set $\mathcal{F} \subset
\mathcal{E}$. Let $D$ be a matrix larger than or equal to
$D_{\{u,v\}}D_{\{u,v\}}^T$ for all the lines $\{u,v\}\in
\mathcal{F}.$ We have the following result for the robust
screening of contingencies. If the inequality
\eqref{eq.BoundingCondition} holds with $D_{\{u,v\}}D_{\{u,v\}}^T$
replaced by $D$, and the clearing time $\tau_{clearing}$ satisfies
$\tau_{clearing}< 2\gamma (V_{\min}-V(x_{pre}))$,  then, for any
contingency associated with the tripping of any line $\{u,v\}\in
\mathcal{F},$ the fault-cleared state $x_F(\tau_{clearing})$ is
still inside the region of attraction of the post-fault SEP
$\delta^*_{post}$, and the post-fault dynamics following the
considered contingency leads to the stable operating condition
$\delta^*_{post}$. This result is a straightforward corollary of
the stability certificate in Section \ref{sec.EstimationCCT}, and
thus its proof is omitted here.

\section{Numerical Illustrations}
\label{sec:simulations}

\subsection{Classical 2-Bus lossy System with Different Pre-fault and Post-fault SEPs}
\begin{figure}[t!]
\centering
\includegraphics[width = 3.2in]{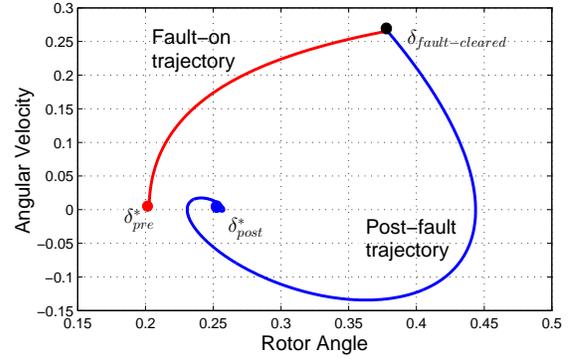}
\caption{System trajectory according to the fault-on dynamics and
post-fault dynamics with the clearing time $CT=2\gamma
(V_{\min}-V(x_{pre}))=1.0600 s$} \label{fig.Trajectory}
\end{figure}

\begin{figure}[t!]
\centering
\includegraphics[width = 3.2in]{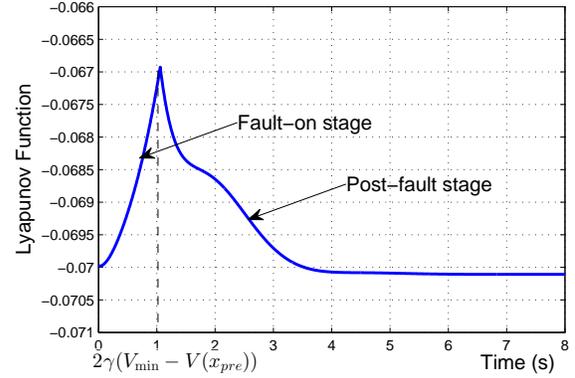}
\caption{Dynamics of the Lyapunov function during the fault-on
stage and post-fault stage with the clearing time $CT=2\gamma
(V_{\min}-V(x_{pre}))=1.0600 s$} \label{fig.Lyapunov}
\end{figure}

For illustrating the presented concepts, this section presents the
simulation results on the most simple 2-bus lossy power system,
described by the single 2-nd order differential equation
\begin{align}
  m \ddot{\delta} +d \dot{\delta} + a \sin(\delta+\alpha) - p=0.
\end{align}
For numerical simulations, we choose $m=0.1$ p.u., $d=0.15$ p.u.,
$a= 0.2$ p.u., and $\alpha=0.05$ rad. The pre-fault and post-fault
power inputs are $p_{pre}=0.05$ p.u. and $p_{post}=0.06$ p.u.
Then, the pre-fault and post-fault stable equilibrium point are
given by $[\delta^*_{pre} \;\;0]^T=[0.2027 \;\; 0]^T$ and
$[\delta^*_{post} \;\;0]^T=[0.2547 \;\; 0]^T,$ both of which are
in the polytope $\Delta(\pi/10).$ Hence,
$\beta=(\sin(\pi/2+\alpha)-\sin(\pi/10+\alpha))/(\pi/2-\pi/10)=0.5114.$
By varying $\gamma$ and solving the LMI
\eqref{eq.BoundingConditionLMI}, we obtain the corresponding lower
bounds for the critical clearing time as in Tab. \ref{tab.CCT}.

\begin{table}[ht!]
\centering
\begin{tabular}{|c|c|}
  \hline
  $\gamma$ & $2\gamma (V_{\min}-V(x_{pre})) (s)$ \\
  \hline
  1 &  0.9442 \\
  2 & 0.9757 \\
  3 & 1.0077 \\
  4 & 1.0297 \\
  5 & 1.0439\\
  6 &  1.0535\\
  7 & 1.0600 \\
  8 &  1.0578 \\
  9 & 1.0574\\
 10 & 1.0553\\
  \hline
\end{tabular}
\caption{Lower bound of the critical clearing time vs.
$\gamma$}\label{tab.CCT}
\end{table}

Therefore, in these values of $\gamma,$ with $\gamma=7$  we obtain
the largest lower bound for the critical clearing time as
$1.0600.$ The corresponding matrices $Q,K,H$ are
\begin{align}
Q=\left[   \begin{array}{ccccc}
          0.0443&    0.0127\\
    0.0127  &  0.0879\\
        \end{array}\right]; K= 0.0968;H= 0.2412,
\end{align}
while the corresponding  value of $V_{\min}-V(x_{pre})$ is
$0.0528.$ In Fig. \ref{fig.Trajectory} we show the dynamics of the
system trajectory in the fault-on and post-fault-stage in which
the clearing time is taken as $\tau_{clearing}=2\gamma
(V_{\min}-V(x_{pre}))=1.0600 s.$ It can be seen that when the
fault happens, the system evolves according to the fault-on
dynamics and the system trajectory deviates from the pre-fault
equilibrium point $\delta^*_{pre}$ to the fault-cleared state
$\delta_{fault-cleared}.$ After the fault self-clears, the system
trajectory recovers from the fault-cleared state
$\delta_{fault-cleared}$ to the post-fault equilibrium point
$\delta^*_{post}$ which is different from the pre-fault
equilibrium. Figure \ref{fig.Lyapunov} shows the divergence of the
Lyapunov function during the fault-on stage and the convergence of
Lyapunov function during the post-fault stage. These figures
confirm the estimation of the critical clearing time as obtained
by the proposed method in this paper.

\subsection{Three Generator System}

Consider the system of three generators with the time-invariant
terminal voltages and mechanical torques given in Tab.
\ref{tab.data3machine}.

\begin{table}[ht!]
\centering
\begin{tabular}{|c|c|c|}
  \hline
  Node & V (p.u.) & $P_k$ (p.u.) \\
  \hline
  1 & 1.0566 & -0.2464 \\
  2 & 1.0502 & 0.2086 \\
  3 & 1.0170 & 0.0378 \\
  \hline
\end{tabular}
\caption{Voltage and mechanical input} \label{tab.data3machine}
\end{table}

The susceptances of the transmission lines are $B_{12}=0.739$
p.u., $B_{13}=1.0958$ p.u., and $B_{23}=1.245$ p.u. The
equilibrium point is calculated as: $\delta^*=[-0.6634\;
   -0.5046\;
   -0.5640 \;0\;0\;0]^T,$ which belongs to the polytope $\Delta(\pi/10).$ Hence, we can take $\beta=(1-\sin(\pi/10))/(\pi/2-\pi/10).$
   For simplicity we just take $m_k=2,d_k=1, k=1,2,3.$ Assume that the fault
   happens at the transmission line connecting
   generators $1$ and $2$ and then self-clears. Also, during that time the mechanical inputs are assumed to be unchanged.
   Taking $\gamma=3$ and using CVX software we can solve the LMI
   \eqref{eq.BoundingConditionLMI} we obtain $Q$ as
   \begin{align}
   \left[%
\begin{array}{cccccc}
    3.8376  &  3.8012 &   3.5779  &  7.5549  &  7.4619  &  7.4166 \\
    3.8012  &  3.8457  &  3.5698  &  7.4776  &  7.5530  &  7.4029 \\
    3.5779  &  3.5698  &  4.0690  &  7.4010  &  7.4185  &  7.6140 \\
    7.5549  &  7.4776  &  7.4010  & 38.9402  & 38.2449  & 38.0704 \\
    7.4619  &  7.5530  &  7.4185  & 38.2449  & 38.9534  & 38.0571 \\
    7.4166  &  7.4029  &  7.6140  & 38.0704  & 38.0571  & 39.1280 \\
    \end{array}%
\right]
   \end{align}
   and $K= \emph{\emph{diag}}(0.2554,
   \;       0.3638,
   \;       0.4386), H=  \emph{\emph{diag}}(0.0943,0.2533,0.2960).$ The corresponding estimation of the critical clearing time is $2\gamma (V_{\min}-V(x_{pre}))= 0.2376 s.$

\subsection{Kundur 9-Bus 3-Generator System}
\begin{figure}[t!]
\centering
\includegraphics[width = 3.2in]{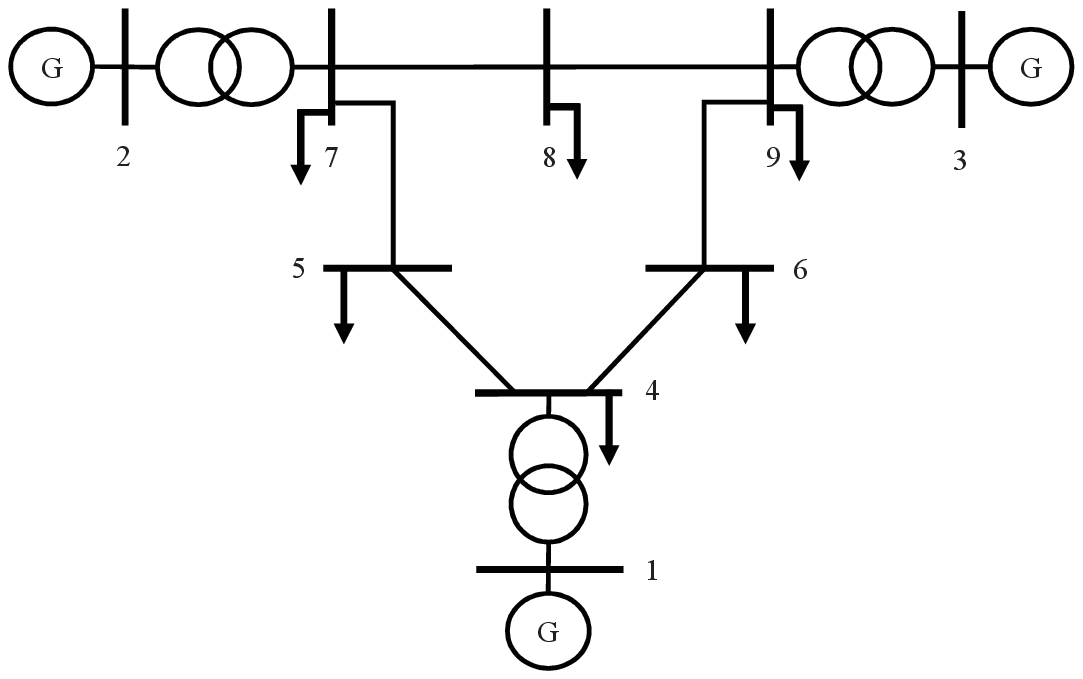}
\caption{3 generator 9 bus system with frequency-dependent dynamic
loads} \label{fig.3generator9bus}
\end{figure}

Consider the Kundur 9 bus 3 machine system depicted in Fig.
\ref{fig.3generator9bus}  with 3 generator buses and 6 load buses.
The susceptances of the transmission lines are as follows:
$B_{14}=17.3611 p.u.,B_{27}=16.0000 p.u.,B_{39}= 17.0648 p.u.,
B_{45}=11.7647 p.u., B_{57}= 6.2112p.u., B_{64}=10.8696p.u.,
B_{78}= 13.8889p.u.,B_{89}=9.9206p.u., B_{96}=5.8824p.u.$ The bus
voltages $V_k$, mechanical inputs $P_{m_k}$, and steady state load
$-P_{d_k}^0$ are given in Tab. \ref{tab.data9bus}. The stable
operating condition is obtained by solving equations
\eqref{eq.operatingCondition} as $x^*=[0.0381\;
    0.3208\;
    0.1924\;
   -0.0349\;
   -0.0421\;
   -0.0409\;
    0.0519\;
    0.0178\;
    0.0155\; 0\; 0\; 0\; 0\; 0\; 0\; 0\; 0\; 0],$ which stays in the polytope $\Delta(\pi/8).$ Hence $\beta=(1-\sin(\pi/8))/(\pi/2-\pi/8)= 0.5240.$ The parameters for generators are $m_1=0.1254, m_2=0.034, m_3=0.016, d_1=0.0627, d_2=0.017, d_3=0.008.$ For simplicity, we take $d_k=0.05, k=4\dots,9.$ Assume that the fault trips the line between buses $6$ and $4$ and when the fault is cleared this line is re-closed. With $\gamma=7.10^{-6}, $ using the CVX software, we can solve the LMI
   \eqref{eq.BoundingConditionLMI} in 1s to obtain the Lyapunov function. Accordingly, we can calculate the minimum value of the Lyapunov function and obtain the estimation for the critical clearing time as $2\gamma (V_{\min}-V(x_{pre}))=0.1175 s.$
\begin{table}[ht!]
\centering
\begin{tabular}{|c|c|c|}
  \hline
  Node & V (p.u.) & $P_k$ (p.u.) \\
  \hline
  1 & 1.0284 & 0.6700 \\
  2 & 1.0085 & 1.6300 \\
  3 & 0.9522 &  0.8500 \\
  4 & 1.0627 & -0.5000 \\
  5 & 1.0707 & -0.7500 \\
  6 & 1.0749 & -0.4500 \\
  7 & 1.0490 & -0.4500 \\
  8 & 1.0579 &  -0.5000 \\
  9 & 1.0521 &  -0.5000 \\
  \hline
\end{tabular}
\caption{Bus voltages, mechanical inputs and static
loads}\label{tab.data9bus}
\end{table}

We perform time domain simulations to find the critical clearing
time for the system when the generators are modeled by swing
equations and by $4^{th}$ orders machine models incorporating
generators' control systems. Accordingly, we can find that when
the fault happens at the transmission line $\{4,6\},$ the true
critical clearing times for the swing model and $4^{th}$ orders
machine models are, respectively, $0.25$s and $0.18$s. Therefore,
the critical clearing time estimated by the proposed method in
this paper is about half of the true one. We conclude that the
proposed method is conservative in comparison to the time domain
simulations, but there is no overestimation for the CCT. In
addition, the time domain simulations confirm the analysis we
described in Remark 1 that the generators' control systems make
the critical clearing time to reduce.

In comparison to the controlling UEP method, the proposed method
in this paper is also more conservative since the controlling UEP
was reported \cite{Chiang:2011eo} to get the estimate for critical
clearing time which is different in less than $10\%$ from the true
one obtained by time domain simulation. However, we note that the
CCT estimate proposed in this paper does not require time-domain
simulation for the fault-on dynamics as in the controlling UEP
method. This will help significantly reduce the computational
resources spent for contingency screening. Therefore, the proposed
framework in this paper can be considered as a complement of the
time domain simulation method and controlling UEP method, which
could be efficiently used when we aim to screen non-critical
contingencies with little computational resources.

\section{Conclusions and Path Forward}
\label{sec:discussion}

In this paper, we introduced techniques to screen contingencies
for transient stability without relying on any time-domain
simulations. This is based on extending the recently introduced
LFF transient stability certificate in the combination with
bounding for the fault-on dynamics. Basically, the LFF approach
can certify the post-fault dynamics's stability when the
fault-cleared state is in some polytope surrounding the post-fault
stable operating point and the Lyapunov function at the
fault-cleared state is under some threshold. We observed that the
LFF certificate only needs to know the fault-cleared state,
instead of the fault-on trajectory. Therefore, with the introduced
bounding technique we can bound the Lyapunov function at the
fault-cleared state, by which we certify stability for a given
contingency scenario without involving any simulations for the
fault-on trajectory and post-fault trajectory. In turns, we
obtained an algebraic formulation for the lower bound of the
critical clearing time, and hence the stability assessment only
involved checking if the clearing time is smaller than that lower
bound to assure the stability of the post-fault dynamics.
Remarkably, the proposed stability certificate only relies on
solving convex optimization problems. It may be therefore scalable
to contingency screening of large scale power systems, especially
when combined with the recent advances in semi-definite
programming exploiting the relatively low tree-width of the grids'
graph \cite{Javadmadani2014sdp}.

Toward the practical applications of the proposed simulation-free
approach to contingency screening, further extensions should be
made in the future where more complicated models of power systems
and faults are considered,
  e.g. generators' control systems, effects of buses' reactive power, and permanent faults are incorporated.
First, since the LFF method is applicable to lossy power grid
\cite{VuTuritsyn:2014acc}, it is possible to extend the proposed
method in this paper to incorporating reactive power, which will
introduce the cosine term in the model \eqref{eq.Bilinear}. This
can be done by extending the state vector $x$ and combining the
technique in this paper with the LFF transient stability
techniques for lossy power grids (without reactive power
considered) \cite{VuTuritsyn:2014acc}. Second, we can see that, in
order to make sure the Lyapunov function is decreasing in the
polytope $\mathcal{Q},$ it is not necessary to restrict the
nonlinear terms $F(Cx)$ to be univariate. As such, we can extend
the proposed method to power systems with generators' voltage
dynamics in which the voltage variable is incorporated in a
multivariable nonlinear function $F.$ Last, the important class of
permanent faults, which will also result in non-identical
pre-fault and post-fault SEPs, should be considered in the future
work with further mathematical development for the representation
of system dynamics under faults and more sophisticated estimation
of critical clearing time.

In the applications, the proposed simulation-free contingency
screening method could be developed to robustly assess the
stability of power systems when a set of faults happen. This will
be applicable to assess major blackout. Also, such a robust
certificate can be applied when there are significant changes in
the power gird topology such as in load shedding
\cite{7077021,siddiqui2015preventive,5706912} and controlled
islanding schemes
\cite{januszquiros2014constrained,6774471,6980139,7024905,pfitzner2011controlled}.
For this end, a more restrictive bounding of the fault-on dynamics
should be employed to alleviate the conservativeness of the
proposed method which is expected when multiple faults are
considered.


\section{Appendix}

\subsection{Proof of the Transient Stability Certificate}
\label{appen.NewQKH}

From  the inequality \eqref{eq.NewQKH}, there exist matrices
$X_{|\mathcal{E}| \times (n+m)}, Y_{|\mathcal{E}|
\times|\mathcal{E}|}$
  such that
\begin{align}
  A^TQ+QA -2\beta C^THC = & -X^TX, \nonumber \\
  QB-(1+\beta)C^TH-(KCA)^T = &-X^TY, \nonumber \\
  -2H =& -Y^TY. \nonumber
\end{align}
The derivative of $V(x)$ along \eqref{eq.Bilinear} is hence given
by:
\begin{align}
    &\dot{V}(x) = 0.5\dot{x}^T Q x+ 0.5x ^T Q\dot{x} \nonumber \\
    &-\sum K_{\{k,j\}}(-\sin\delta_{{kj}}+\sin\delta_{kj}^*)\dot{\delta}_{{kj}}
    \nonumber \\ &=0.5x^T(A^TQ+QA)x-x^TQBF   + F^TKC\dot{x} \nonumber \\
    &=0.5x^T(2\beta C^THC-X^TX)x  \nonumber \\ &- x^T\big((1+\beta)C^TH+(KCA)^T-X^TY\big)F
     \nonumber \\ &+ F^TKC(Ax-BF)
\end{align}
Noting that $CB=0$ and $Y^TY=2H$ yields
\begin{align}
 &\dot{V}(x)=-0.5(Xx-YF)^T(Xx-YF)    + \sum H_{\{k,j\}}g_{\{k,j\}} \nonumber \\
  \end{align}
where
$g_{\{k,j\}}=(f_{\{k,j\}}-(\delta_{kj}-\delta_{kj}^*))(f_{\{k,j\}}-\beta(\delta_{kj}-\delta_{kj}^*))
\le 0, \forall x\in \mathcal{Q}.$ As such, the Lyapunov function
$V(x)$ is decaying inside the polytope $\mathcal{Q}.$ The other
results immediately follow those in \cite{Vu:2014}.

\subsection{Proof of the Bounding of Fault-on Dynamics}
\label{appendix.BoundingCondition}

From  the inequality \eqref{eq.BoundingCondition}, there exist
matrices $X_{|\mathcal{E}| \times (n+m)}, Y_{|\mathcal{E}|
\times|\mathcal{E}|}$
  such that
\begin{align}
  A^TQ+QA -2\beta C^THC+ \gamma (QBD_{uv})(QBD_{uv})^T = & -X^TX, \nonumber \\
  QB-(1+\beta)C^TH-(KCA)^T = &-X^TY, \nonumber \\
  -2H =& -Y^TY. \nonumber
\end{align}
Similar to the above section, we obtain
\begin{align}
\label{eq.dotV} &\dot{V}(x_F)=-0.5(Xx_F-YF_{pre})^T(Xx_F-YF_{pre})    + \sum H_{\{k,j\}}g_{\{k,j\}_F} \nonumber \\
    & +x_F^TQBD_{\{u,v\}}\sin\delta_{uv_F}-0.5\gamma x_F^T(QBD_{\{u,v\}})(QBD_{\{u,v\}})^Tx_F
  \end{align}
where
$g_{\{k,j\}_F}=(f_{\{k,j\}}-(\delta_{kj_F}-\delta_{{kj}_{pre}}^*))(f_{\{k,j\}}-\beta(\delta_{kj_F}-\delta_{kj_{pre}}^*)).$

 Note that
\begin{align}
 g_{\{k,j\}_F} \le &0, \forall x_F \in \mathcal{Q}, \forall \delta^*_{pre}\in\Delta(\lambda), \nonumber \\
 x_F^TQBD_{\{u,v\}}\sin\delta_{uv_F} \le & 0.5\gamma x_F^T(QBD_{\{u,v\}})(QBD_{\{u,v\}})^Tx_F \nonumber \\
 &+ 0.5\sin^2\delta_{uv_F}/\gamma \nonumber \\
                                     \le & 0.5\gamma x_F^T(QBD_{\{u,v\}})(QBD_{\{u,v\}})^Tx_F \nonumber \\&+ \frac{1}{2\gamma}.
\end{align}

Hence, $\dot{V}(x_F) \le \dfrac{1}{2\gamma}$ whenever $x_F \in
\mathcal{Q}.$

\subsection{Proof of The Clearing Time-based Stability Certificate}
\label{appendix.ClearingTimeCertificate} We will prove that with
$\tau_{clearing}<2\gamma(V_{\min}-V(x_{pre})),$ the fault-cleared
state $x_F(\tau_{clearing})$ is still in the set $\mathcal{R_Q}.$

Note that the boundary of the set $\mathcal{R_Q}$ is composed of
segments which belong to sublevel set of the Lyapunov function
$V(x)$ and segments which belong to the flow-in boundaries
$\partial\mathcal{Q}^{in}_{kj}$ which is defined by
$|\delta_{kj}|=\pi/2$ and $\delta_{kj}\dot{\delta}_{kj}<0.$ It is
easy to see that the flow-in boundaries
$\partial\mathcal{Q}^{in}_{kj}$ prevent the fault-on dynamics
\eqref{eq.fault-on} from escaping $\mathcal{R_Q}.$

Assume that $x_F(\tau_{clearing})$ is not in the set
$\mathcal{R_Q}.$ Then the fault-on trajectory can only escape
$\mathcal{R_Q}$ through the segments which belong to sublevel set
of the Lyapunov function $V(x).$ Denote $\tau$ be the first time
at which the fault-on trajectory meets one of the boundary
segments which belong to sublevel set of the Lyapunov function
$V(x).$ Hence $x_F(t) \in \mathcal{R_Q}$ for all $0 \le t \le
\tau.$ Since $\dot{V}(x_F) \le \dfrac{1}{2\gamma}$ whenever $x_F
\in \mathcal{Q},$ and the fact that $\mathcal{R_Q}\subset
\mathcal{Q},$ we have

\begin{align}
V(x_F(\tau))-V(x_F(0)) = \int_0^{\tau} \dot{V}(x_F(t))dt \le
\frac{\tau}{2\gamma}
\end{align}
Hence $\tau \ge 2\gamma
(V(x_F(\tau))-V(x_F(0)))=2\gamma(V(x_F(\tau))-V(x_{pre})).$ By
definition of $\tau$, we have $V(x_F(\tau))=V_{\min}.$ Therefore,
$\tau \ge 2\gamma(V_{\min}-V(x_{pre}))$ and thus
$\tau_{clearing}\ge 2\gamma(V_{\min}-V(x_{pre})),$ which is a
contradiction.

\section{Acknowledgements}
This work was partially supported by  Masdar, MIT/Skoltech
initiatives, and Ministry of Education and Science of Russian
Federation, Grant Agreement no. 14.615.21.0001. We thank the
anonymous reviewers for their careful reading of our manuscript
and their many valuable comments and constructive suggestions
which helped to significantly improve the quality of this paper.

\bibliographystyle{IEEEtran}
\bibliography{lff}
\end{document}